\documentclass[twocolumn]{aastex62}

\graphicspath{{./}{figures/}}

\received{2018-08-15}
\revised{2018-09-13}
\accepted{2018-09-17}
\submitjournal{ApJL}

\shorttitle{SOFIA detects spatially resolved [\ion{C}{2}] emission in HE0433-1028}
\shortauthors{Busch et al.}

\begin{document}

\title{The Close AGN Reference Survey (CARS):\\SOFIA detects spatially-resolved [\ion{C}{2}] emission in the luminous AGN HE0433-1028\footnote{Based on observations made with the NASA/DLR Stratospheric Observatory for Infrared Astronomy (SOFIA) and on observations collected at the European Organisation for Astronomical Research in the Southern Hemisphere under ESO programme 094.B-0354(A).}}

\correspondingauthor{Gerold Busch, Bernd Husemann}
\email{busch@ph1.uni-koeln.de, husemann@mpia.de}

\author[0000-0001-6679-5481]{G. Busch}
\affil{I.~Physikalisches Institut der Universit\"at zu K\"oln, Z\"ulpicher Str. 77, 50937 K\"oln, Germany}

\author[0000-0003-2901-6842]{B. Husemann}
\affiliation{Max-Planck-Institut f\"ur Astronomie, K\"onigstuhl 17, 69117 Heidelberg, Germany}

\author[0000-0002-2260-3043]{I. Smirnova-Pinchukova}
\affiliation{Max-Planck-Institut f\"ur Astronomie, K\"onigstuhl 17, 69117 Heidelberg, Germany}

\author[0000-0001-6049-3132]{A. Eckart}
\affil{I.~Physikalisches Institut der Universit\"at zu K\"oln, Z\"ulpicher Str. 77, 50937 K\"oln, Germany}
\affil{Max-Planck-Institut f\"ur Radioastronomie, Auf dem H\"ugel 69, 53121 Bonn, Germany}

\author[0000-0002-4735-8224]{S. A. Baum}
\affil{Carlson Center for Imaging Science,  Rochester Institute of Technology, 84 Lomb Memorial Drive, Rochester, NY 14623, USA}
\affil{Faculty of Science, University of Manitoba, Winnipeg, MB R3T 2N2, Canada}

\author[0000-0003-2658-7893]{F. Combes}
\affil{LERMA, Observatoire de Paris, PSL Research Univ., Coll\`ege de France, CNRS, Sorbonne Univ., UPMC, Paris, France}

\author[0000-0003-2880-9197]{S. M. Croom}
\affil{Sydney Institute for Astronomy, School of Physics, University of Sydney, NSW 2006, Australia}

\author[0000-0003-4932-9379]{T. A. Davis}
\affil{School of Physics \& Astronomy, Cardiff University, Queens Buildings, The Parade, Cardiff, CF24 3AA, UK}

\author[0000-0001-7163-536X]{N. Fazeli}
\affil{I.~Physikalisches Institut der Universit\"at zu K\"oln, Z\"ulpicher Str. 77, 50937 K\"oln, Germany}

\author[0000-0003-2649-3707]{C. Fischer}
\affil{Deutsches SOFIA Institut, Pfaffenwaldring 29, 70569 Stuttgart, Germany}

\author[0000-0003-2754-9258]{M. Gaspari}
\altaffiliation{Einstein and Spitzer fellow}
\affil{Department of Astrophysical Sciences, Princeton University, 4 Ivy Lane, Princeton, NJ 08544-1001, USA}

\author[0000-0002-7187-9126]{R. Klein}
\affiliation{SOFIA/USRA, NASA Ames Research Center, Moffett Field, California 94035, USA}

\author{M. Krumpe}
\affil{Leibniz-Institut f\"ur Astrophysik Potsdam, An der Sternwarte 16, 14482 Potsdam, Germany}

\author[0000-0003-2290-7060]{R. McElroy}
\affil{Max-Planck-Institut f\"ur Astronomie, K\"onigstuhl 17, 69117 Heidelberg, Germany}

\author[0000-0001-6421-054X]{C. P. O'Dea}
\affil{Department of Physics \& Astronomy, University of Manitoba, Winnipeg, MB R3T 2N2, Canada}
\affil{School of Physics \& Astronomy, Rochester Institute of Technology, 84 Lomb Memorial Drive, Rochester, NY 14623, USA}

\author[0000-0001-5654-0266]{M. A. Perez-Torres}
\affil{Instituto de Astrof\'isica de Andaluc\'ia - Consejo Superior de
Investigaciones Cient\'ificas (CSIC), PO Box 3004, 18008, Granada, Spain}
\affil{Departamento de F\'isica Te\'orica, Facultad de Ciencias, Universidad de Zaragoza, 50019, Zaragoza, Spain}

\author{M. C. Powell}
\affil{Yale Center for Astronomy and Astrophysics, Yale University, PO Box 2018120, New Haven, CT 06520-8120, USA}

\author[0000-0002-3078-9482]{\'A. S\'anchez-Monge}
\affil{I.~Physikalisches Institut der Universit\"at zu K\"oln, Z\"ulpicher Str. 77, 50937 K\"oln, Germany}

\author[0000-0003-1585-9486]{J. Scharw\"achter}
\affil{Gemini Observatory, Northern Operations Center, 670 N. A’ohoku Place, Hilo, HI 96720, USA}

\author[0000-0002-5445-5401]{G. R. Tremblay}
\affil{Harvard-Smithsonian Center for Astrophysics, 60 Garden St., Cambridge, MA 02138, USA}

\author[0000-0001-6746-9936]{T. Urrutia}
\affil{Leibniz-Institut f\"ur Astrophysik Potsdam, An der Sternwarte 16, 14482 Potsdam, Germany}



\begin{abstract}

We report\deleted{the first} spatially-resolved [\ion{C}{2}]$\lambda 158\,\mu\mathrm{m}$ observations of\deleted{a nearby luminous AGN,} HE 0433--1028, \deleted{at redshift $0.0355$ obtained} \added{which is the first detection of a nearby luminous AGN (redshift 0.0355)} with FIFI-LS onboard the airborne observatory SOFIA. 
We compare the spatially-resolved star formation tracers [\ion{C}{2}], as provided by our SOFIA observations, and H$\alpha$ from MUSE optical integral-field spectroscopy.
We find that the [\ion{C}{2}] emission is mainly matching the extended star formation as traced by the extinction-corrected H$\alpha$ line emission but some additional flux is present. 
While a larger sample is needed to statistically confirm our findings and investigate possible dependencies on AGN luminosity and star formation rate, our study underlines the necessity of collecting a spatially-resolved optical-FIR dataset for nearby AGNs, and shows that it is technically feasible to collect such datasets with FIFI-LS onboard SOFIA.

\end{abstract}



\section{Introduction} \label{sec:intro}

The [\ion{C}{2}] fine-structure line of singly ionized carbon at $157.74\,\mu\mathrm{m}$ is one of the brightest emission lines in the far-infrared and a main cooling line in the interstellar medium and photon-dominated regions \citep[PDR,][]{1999RvMP...71..173H}. Many studies show a correlation between the [\ion{C}{2}] line and the star formation rate (SFR) in nearby galaxies \citep[e.g.][]{1991ApJ...373..423S,2002A&A...385..454B,2014A&A...568A..62D,2015ApJ...800....1H}. While the optical hydrogen recombination line H$\alpha$, another popular star formation rate tracer, is shifted out of the near-infrared at $z\gtrsim 2$,  the [\ion{C}{2}] line shifts to the sub-mm at higher redshift ($z\gtrsim 4$\added{, and even $z\gtrsim 1.5$ in ALMA band 10}) where it can be observed with large interferometers like ALMA or NOEMA at unprecedented sensitivity and resolution, therefore becoming a powerful diagnostic for star formation in high-redshift galaxies \citep[e.g.][]{2010ApJ...724..957S}. 

Although the correlation between the [\ion{C}{2}] line and the SFR \added{for normal star-forming galaxies} is relatively tight, some scatter exists \citep[\added{$\sim 0.2\,\mathrm{dex}$,}][]{2015ApJ...800....1H} and indicates a mix of different excitation mechanisms in individual galaxies. Of particular interest is which influence the active galactic nucleus (AGN) has on the [\ion{C}{2}] luminosity: Luminous AGNs have a strong radiation field that is in some cases able to dominate the ionization of the ISM even out to kiloparsec scales \citep[e.g.][]{2011AJ....142...43S,2013A&A...549A..43H} and could therefore contribute to the excitation of C$^+$ as well. On the other hand, very luminous infrared sources seem to have relatively weaker [\ion{C}{2}] luminosities (``\ion{C}{2} deficit''). \replaced{probably due to}{It is still an open question what is the main driver of this \ion{C}{2} deficit \citep[e.g.][]{2017ApJ...834....5S}. Possible explanations for a \ion{C}{2} deficit to occur close to the AGN include} dust heating by the AGN \citep[e.g.][]{2003ApJ...594..758L,2011ApJ...728L...7G} or carbon overionization due to X-rays \citep[e.g.][]{2015A&A...580A...5L,2018ApJ...861...95H}. 
A reliable baseline needs to be established at low redshift. \replaced{In particular,}{While some nearby low-luminosity AGN are included in e.g. the KINGFISH survey \citep{2015ApJ...800....1H,2017ApJ...834....5S},} the influence of \added{luminous} \replaced{active galactic nuclei}{AGN,} \added{as they are ubiquitous at higher redshift,} on the calibration of the [\ion{C}{2}]-SFR relation needs to be investigated.

Spatially and spectrally resolved analysis is a powerful way to distinguish between \replaced{the different components}{extended and central point source emission} to estimate the impact of the galactic nucleus. However, this kind of study has not yet been attempted due to the lack of high-spatial and -spectral resolution observations of reference star formation indicators. The \emph{Close AGN Reference Survey} \citep[CARS, \url{www.cars-survey.org},][]{2017Msngr.169...42H} comprises a spatially-resolved multiwavelength dataset of nearby ($0.01 < z < 0.06$) luminous AGNs that are ideal as a bridge between the well-studied local AGN population and more distant quasi-stellar objects \citep[QSOs,][]{2007A&A...470..571B,2016A&A...587A.138B}. It therefore provides a reference data set for high-redshift galaxies and also offers a range of AGN luminosities ($L_\mathrm{bol}\sim 10^{43} - 10^{45.5}\,\mathrm{erg}\,\mathrm{s}^{-1}$) that allows for a statistical analysis of possible deviations from the [\ion{C}{2}]-SFR relation as a function of the AGN luminosity.

In this letter, we report the first spatially-resolved detection of [\ion{C}{2}] line emission in a nearby luminous AGN with the airborne \emph{Stratospheric Observatory For Infrared Astronomy} \citep[SOFIA,][]{2012ApJ...749L..17Y}. We test the [\ion{C}{2}]-SFR relation in the presence of a bright AGN by comparing spatially-resolved [\ion{C}{2}] emission with respect to H$\alpha$ as a reference SFR indicator. HE 0433--1028 is a strongly barred spiral galaxy at a redshift of $z=0.0355$ and an AGN luminosity of $4\times 10^{44}\,\mathrm{erg}\,\mathrm{s}^{-1}$. Adopting a cosmology with $H_0 = 70\,\mathrm{km}\,\mathrm{s}^{-1}\,\mathrm{Mpc}^{-1}$, $\Omega_\mathrm{M} = 0.3$ and $\Omega_\Lambda = 0.7$, the redshift corresponds to a luminosity distance of $D_L = 157.4\,\mathrm{Mpc}$ and a scale of $0.707\,\mathrm{kpc}/\arcsec$.

\section{Observations}
\label{sec:obs}

\subsection{FIFI-LS far-infrared 3D spectroscopy}
\label{sec:fifi}

The \emph{Field-Imaging Far-Infrared Line Spectrometer} \citep[FIFI-LS,][]{2010SPIE.7735E..1TK,Fischer2018} onboard the flying telescope SOFIA is an integral-field spectrograph working at far-infrared wavelengths. The AGN HE 0433-1028 was observed on 2016 March 1 at a pressure altitude of ${39000}$ feet, with an on-source total exposure time of $25.6\,\mathrm{min}$. We used the red channel centered on the [\ion{C}{2}]$\lambda 157.7\,\mu\mathrm{m}$ line (redshifted with $z=0.0355$), providing a spectral coverage from $162.8\,\mu\mathrm{m}$ to $163.9\,\mu\mathrm{m}$. In this mode, the array consists of 5$\times$5 pixels with a pixel size of $12\arcsec \times 12\arcsec$ each. The spectral resolution is $R\approx 1200$ or $250\,\mathrm{km}\,\mathrm{s}^{-1}$ at this wavelength.

Chop-subtraction, flat correction, telluric correction, flux calibration and spectral rebinning is performed by the instrument pipeline. Before spatial rebinning, we calculate and subtract the remaining background in each exposure and spatial position by calculating a weighted average of the spectral region around the expected emission line. In this process we also calculate the standard deviation. In the next step, we replace NaN-values in the spectrum with white noise with the previously calculated standard deviation. This ensures a constant noise over the whole spectral range. The spatial alignment and rebinning to a pixel size of $12\arcsec \times 12\arcsec$ is done with a Drizzle algorithm \citep{2002PASP..114..144F}. 

While coadding the drizzled cubes, we realized that not all cubes have the same quality. We therefore decided to observe how the S/N of the line detection behaves while adding more cubes. It became apparent that only during the second half of the observation, the increase of S/N with cumulative exposure time $t_\mathrm{exp}$ followed the expected relation $\mathrm{S/N} \propto \sqrt{t_\mathrm{exp}}$. During the first half of the observations, the S/N stayed constant over time on a low level. We therefore concluded that the cubes observed during the first half were not useful due to technical or weather influence (higher line-of-sight water vapour forecasted at the beginning of flight leg) and only coadded the second half. This left us with 25 accepted and 25 rejected frames, resulting in an effective exposure time of about $13\,\mathrm{min}$.

\begin{figure}
\centering
\includegraphics[width=\columnwidth]{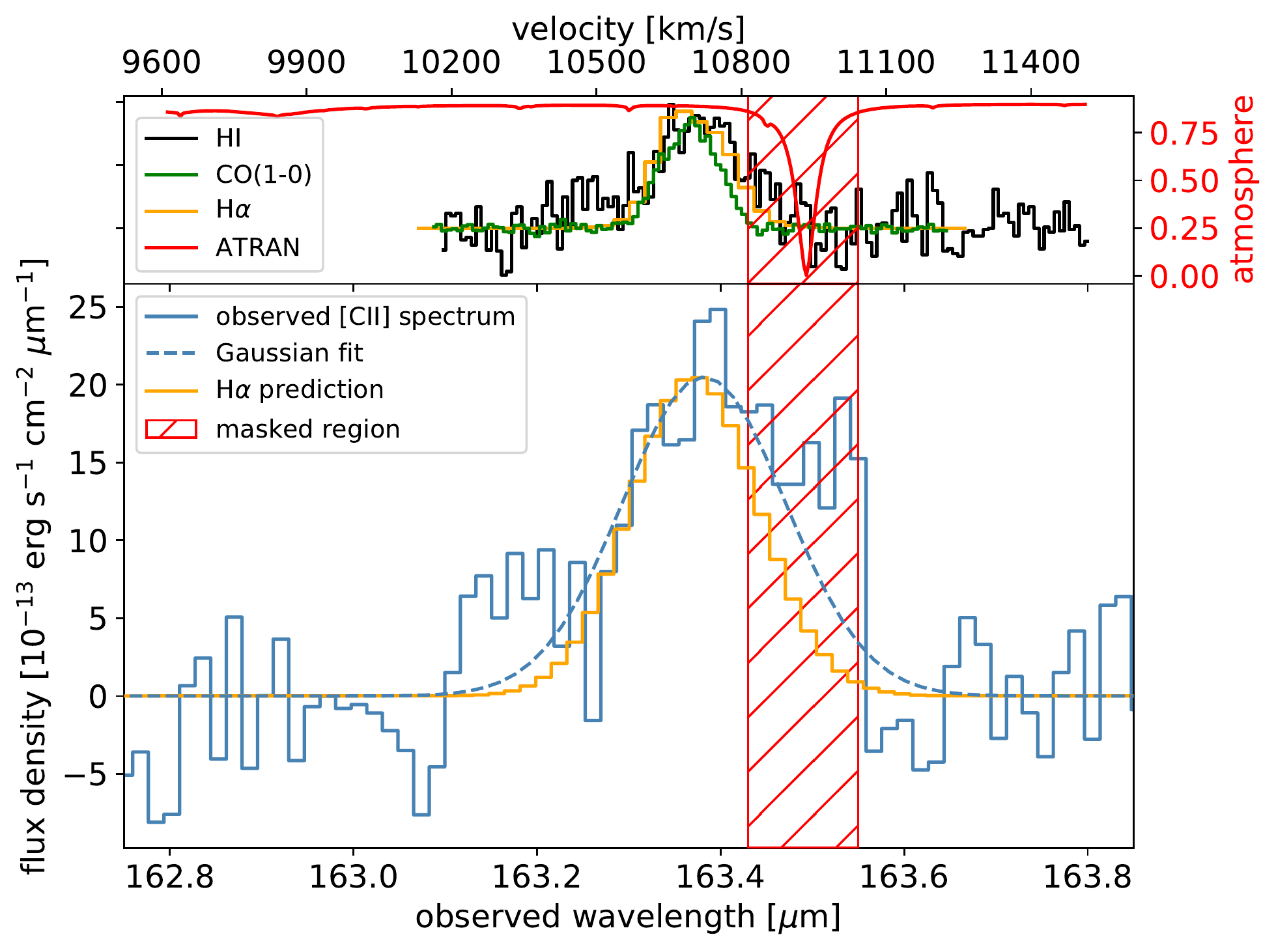}
\caption{\emph{Bottom panel:} FIFI-LS spectrum integrated in an aperture with radius $18\arcsec$ (blue line), approximately corresponding to the FWHM of the PSF. A region masked in the fit because of low atmospheric transmission is marked with red stripes. A Gaussian fit of the [\ion{C}{2}] line, which has a signal-to-noise of $\sim$$18$, is shown in dashed lines. The orange spectrum shows the expected line profile, based on H$\alpha$ observations. \emph{Top panel:} Integrated \ion{H}{1}, CO(1-0) and H$\alpha$ line profile. A synthetic spectrum of the atmospheric transmission from \textsc{Atran} is overplotted in red. It is apparent that no emission is expected in the spectral region with low atmospheric transmission.}
\label{fig:comparison}
\end{figure}

Figure \ref{fig:comparison} (bottom panel, blue line) shows the resulting FIFI-LS spectrum integrated in the central aperture with radius $18\arcsec$, which roughly corresponds to the full width at half maximum (FWHM) of the point-spread function (PSF) of the SOFIA telescope. In orange, we show the scaled H$\alpha$ line (see below), \deleted{adjusted to match the spatial and spectral resolution of SOFIA and} integrated over the same aperture. \added{Beforehand, the MUSE H$\alpha$ datacube was smoothed and regridded to match the spatial resolution of SOFIA and spectral resolution of FIFI-LS (see also Sect.~\ref{sec:model}).}
In the top panel, we show the line profiles of \ion{H}{1} \citep[Nan\c{c}ay radio telescope,][]{2004A&A...416..515D}, CO(1-0) (ALMA, T. Davis, priv. comm.) and H$\alpha$, and overlay in red a synthetic spectrum of the atmospheric transmission created with \textsc{Atran}\footnote{https://atran.sofia.usra.edu/} \citep{1992nstc.rept.....L}. \added{The CO and H$\alpha$ spectra were extracted over the same area as the [\ion{C}{2}] observation, while the \ion{H}{1} observation had a much larger beam of several arcmin.}
We see that there is a strong absorption feature at $163.5\,\mu\mathrm{m}$ but we do not expect any emission in neutral and ionized gas at the velocity of this feature. 
However, this feature might cause strong residuals in the [\ion{C}{2}] emission. We therefore decide to mask this spectral region (red stripes in Fig.~\ref{fig:comparison}) before fitting a Gaussian function that we show as a dashed line in the bottom panel of Fig.~\ref{fig:comparison}. The fitted [\ion{C}{2}] line has a signal-to-noise of $\sim$$18$. The line width is FWHM$\approx 380\,\mathrm{km}\,\mathrm{s}^{-1}$ ($290\,\mathrm{km}\,\mathrm{s}^{-1}$ after correcting for instrumental resolution) and the peak velocity of the [\ion{C}{2}] emission coincides with that of the H$\alpha$ line within $\lesssim 30\,\mathrm{km}\,\mathrm{s}^{-1}$. This demonstrates FIFI-LS ability to detect extragalactic sources, and extends SOFIA's scope from sources in the immediate Galactic neighborhood to nearby galaxies and AGNs.

\subsection{MUSE optical 3D spectroscopy}
\label{sec:obs-muse}

Optical integral-field spectroscopy with the \emph{Multi-Unit Spectroscopic Explorer} \citep[MUSE;][]{2010SPIE.7735E..08B} is available from CARS. HE 0433-1028 was observed in December 2014 with a total on-source integration time of 600 seconds. The reduction with the standard ESO pipeline \cite{2012SPIE.8451E..0BW} results in a data cube with a large field-of-view (FOV) of $1\arcmin \times 1\arcmin$ which matches the FIFI-LS FOV. The spatial scaling is $0\farcs 2$ per spaxel (spatial pixel) and the wavelength coverage is $4750-9300\,$\r{A} with a spectral resolution of $R=2500$ ($\sim$$110\,\mathrm{km}\,\mathrm{s}^{-1}$) at the wavelength of the H$\alpha$ line.

\begin{figure*}
\centering
\includegraphics[width=0.32\linewidth]{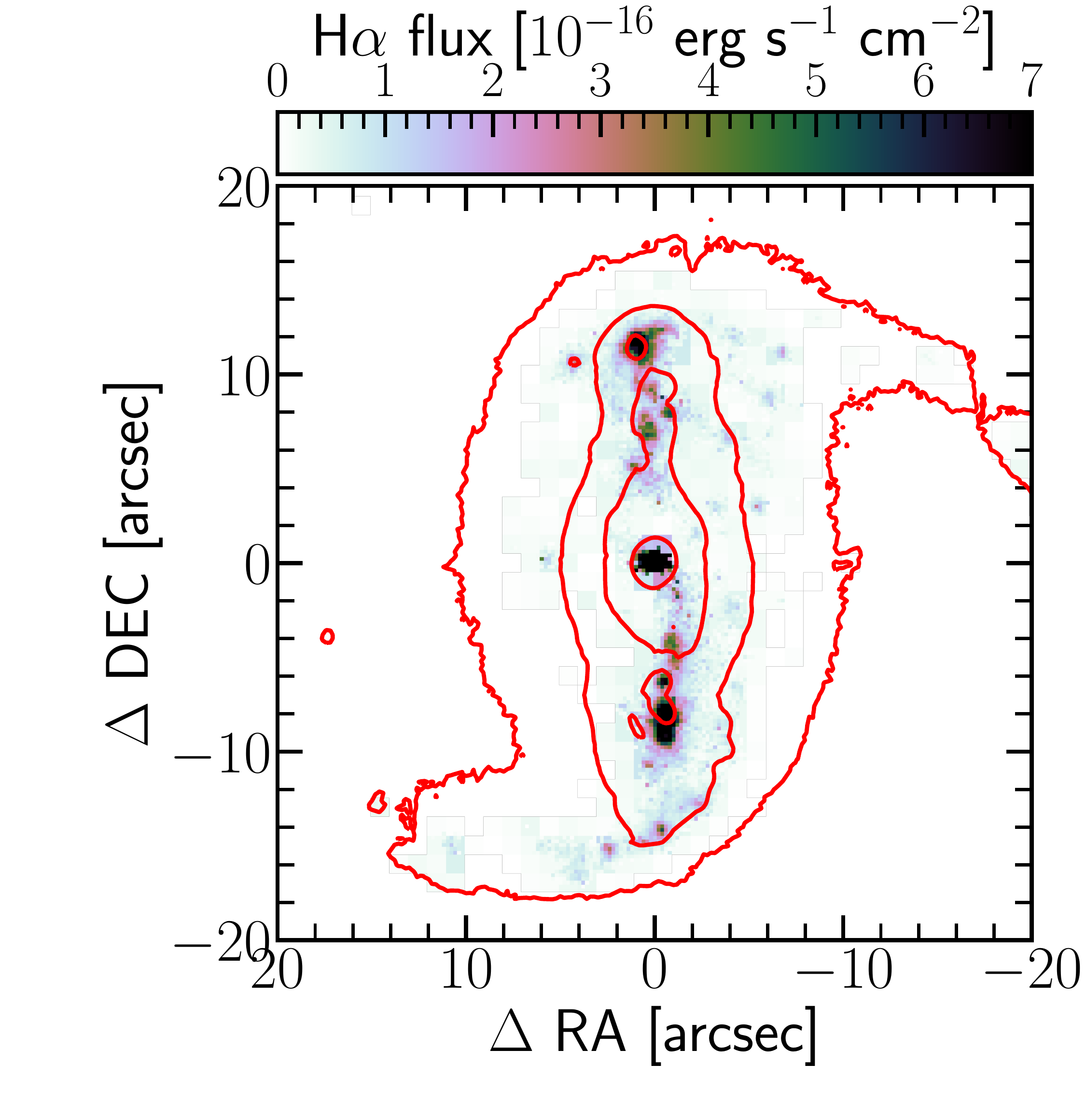}
\includegraphics[width=0.32\linewidth]{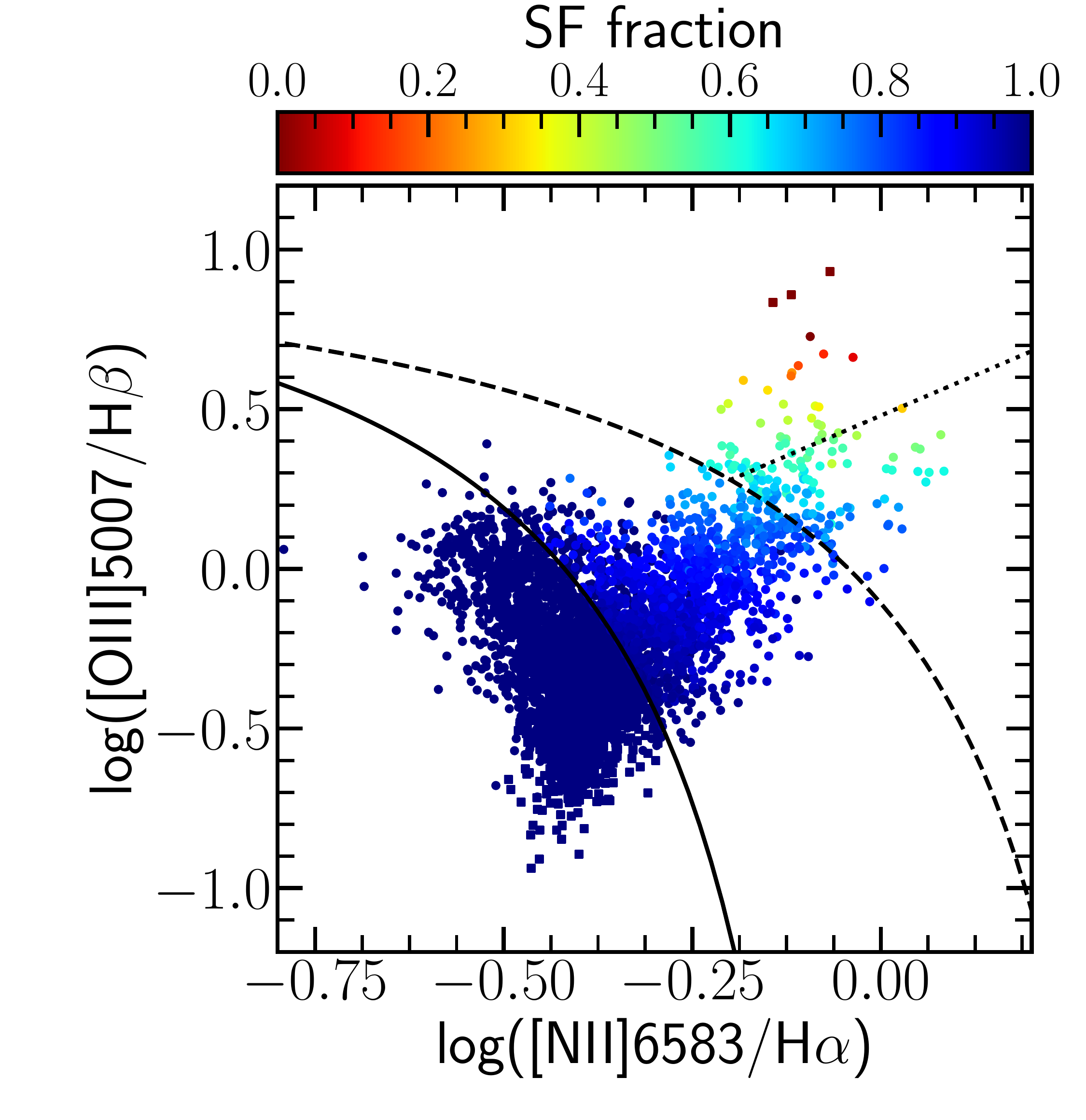}
\includegraphics[width=0.32\linewidth]{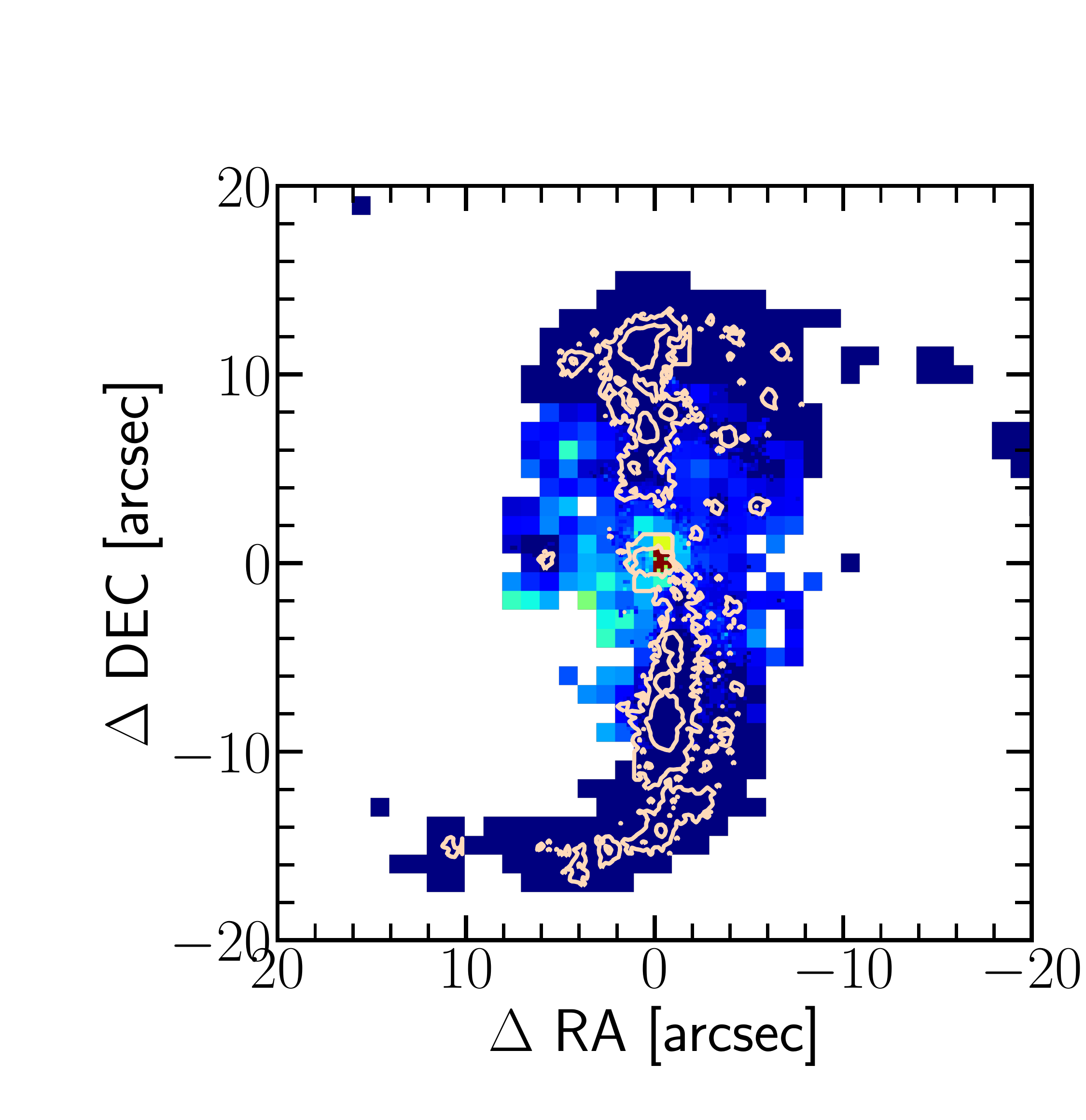}
\caption{\emph{Left:} AGN subtracted and extinction corrected H$\alpha$ flux map, that traces the star formation in the galaxy with high resolution. $i$-band continuum emission is overlayed in red contours. The H$\alpha$ emission shows an elongated distribution since the star formation regions are mostly situated within the stellar bar. \emph{Middle:} Optical BPT-diagram \citep[with division lines from][]{ 2001ApJ...556..121K,2003MNRAS.346.1055K,2010MNRAS.403.1036C} showing the ionization mechanisms of the nebular gas on a spaxel-to-spaxel basis. The AGN fraction is color-coded with red being purely excited by AGN and blue purely by star formation. \emph{Right:} Map of the AGN fraction, with H$\alpha$ contours overlayed.
The flux in each pixel of {\tt HaMap} is lowered by the respective AGN fraction to resemble the extended emission based on star formation.}
\label{fig:haamap}
\end{figure*}

The tool QDeblend$^{\mathrm{3D}}$ makes use of the fact that HE 0433-1028 is an unobscured AGN and the broad-line region appears point-like. By combining spatial and spectral information, it is possible to deblend central point-like and extended emission in an iterative process \citep{2014MNRAS.443..755H}. In a next step, the stellar continuum was subtracted and optical emission lines were fitted with Gaussian functions across the cube. Using the Balmer decrement H$\alpha$/H$\beta$, extinction correction was then performed following the extinction law of \cite{1989ApJ...345..245C}. Figure \ref{fig:haamap} (left) shows a map of the extinction-corrected H$\alpha$ extended emission with the $i$-band continuum emission overlayed as contours.

Afterwards we analyze the excitation mechanisms of the ionized gas across the galaxy. We show the BPT diagram \citep{1981PASP...93....5B}, which uses the ratios of nearby optical emission lines to distinguish the ionization mechanism of nebular gas, in Fig.~\ref{fig:haamap}, middle panel. \added{In the diagram, we fit a straight mixing line in linear-linear space. The \ion{H}{2} and narrow-line region AGN basis spectra are the ones with line ratios closest to the minimal and maximal points of this line. Assuming that all spectra are a linear combination of the \ion{H}{2} and AGN basis spectra, we can now derive the AGN and star forming fraction \citep{2016MNRAS.462.1616D}.} The color-coding shows the AGN fraction \deleted{from the AGN-SF mixing sequence in the BPT diagram, derived} for each pixel. The right panel shows the AGN fractions in the map. 
The H$\alpha$ flux in each pixel is then lowered by the respective AGN fraction. This leaves us with an AGN subtracted H$\alpha$ map which is representative of the extended star formation.

\section{Modeling the [\ion{C}{2}] emission}
\label{sec:model}
\begin{figure*}
\centering
\includegraphics[width=\linewidth]{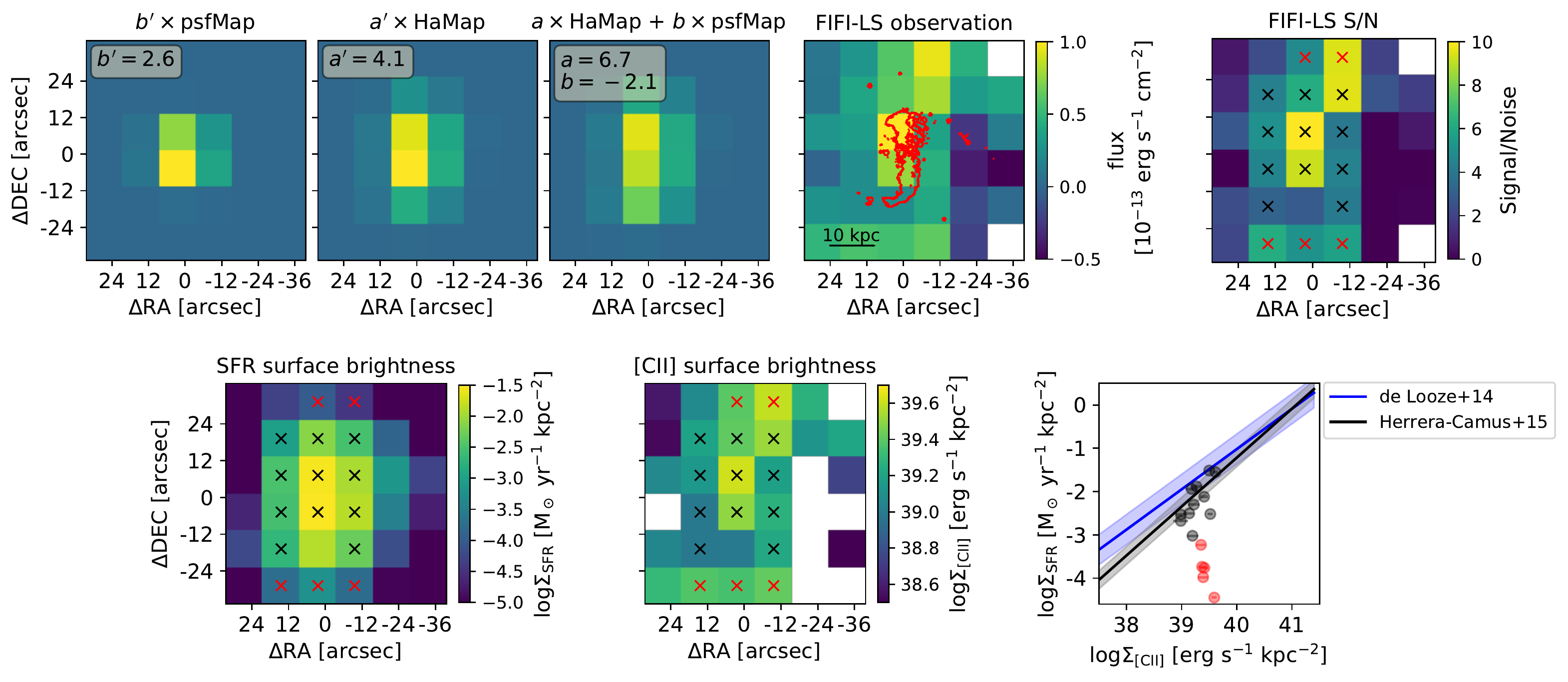}
\caption{\emph{From left to right\added{, top row}:} We compare three different models, only point source, only extended emission and a combined model, each fitted to the observed [\ion{C}{2}] line map from FIFI-LS obtained by integrating over the wavelength interval from $163.2\,\mu \mathrm{m}$ to $163.43\,\mu \mathrm{m}$. It becomes apparent that the line distribution is extended. In the fourth panel we show the observed [\ion{C}{2}] emission line map with contours of H$\alpha$ from the MUSE cube overlayed in red. In the last panel we show the signal-to-noise map of our observations\replaced{which indicates that the central pixels have the highest signal-to-noise ratio}{. Pixels with S/N$>3$ are marked with crosses. \emph{Bottom row:} Surface brightnesses of star formation rate and [\ion{C}{2}] emission. Spatially-resolved [\ion{C}{2}]-SFR relation, with relations from the literature indicated.}}
\label{fig:maps}
\end{figure*}

With the available FIFI-LS data we can spatially distinguish between the central region that might be influenced by the AGN and the star formation along the bar.
In Fig.~\ref{fig:maps}, fourth panel, we show the flux distribution of the [\ion{C}{2}] line, which we derive by summing up the spectra on a spaxel-by-spaxel basis between $163.2\,\mu\mathrm{m}$ and $163.43\,\mu\mathrm{m}$. The \added{first value was chosen based on the H$\alpha$ spectrum and the fit described in Sect.~\ref{sec:fifi}\footnote{\added{We confirmed that changing the starting value to e.g. $163.1\,\mu\mathrm{m}$ leads to the same results within the uncertainties.}}, the} latter value was chosen to avoid the atmospheric absorption feature (see Fig.~\ref{fig:comparison}). 
We correct for missing flux in this wing by applying a correction factor of $1.5$ which was derived from the integrated spectrum by comparing the total flux in the Gaussian with the flux derived when summing only in the mentioned interval. 
The resulting map resembles the elongated structure of the star forming regions in the bar, as visible in the H$\alpha$ map in Fig.~\ref{fig:haamap} (left), which has a length of about $30\arcsec$ ($\sim$$20\,\mathrm{kpc}$). \replaced{To quantify this finding, we create a model of the [\ion{C}{2}] emission, which }{When considering only pixels with S/N$>3$, we find an extent of almost $1\arcmin$. However, the length is affected by beam smearing. To further further quantify this finding, we aim at distinguishing between extended emission and unresolved point-source contribution. However, given the small extent of the galaxy ($\sim$$30\arcsec$) compared to the spatial resolution of SOFIA ($\sim$$13\arcsec$), a direct decomposition seems inadvisable. Instead, we use a forward modelling approach with the MUSE H$\alpha$ map as a high spatial-resolution prior, which is also not affected by beam-smearing. Our model }consists of two components: extended emission and an unresolved point source. 

For the extended emission, {\tt HaMap}, we use the extinction-corrected H$\alpha$ flux map (see Fig.~\ref{fig:haamap}) as a prior for the expected [\ion{C}{2}] assuming that both trace the gas excitation from star formation. We convolve the H$\alpha$ map  with the spatial resolution of SOFIA's 2.7m-telescope (PSF approximated with a Gaussian with FWHM$\approx 12\farcs5$ at $163.4\,\mu\mathrm{m}$) and  apply a spatial rebinning to $12\arcsec$  pixel grid as implied by the SOFIA data with respect to the target coordinates. We stress that the AGN-host galaxy deblending step as described in Sect.~\ref{sec:obs-muse}  is essential for using the H$\alpha$ maps as a robust star formation tracer.

The spatially unresolved component, {\tt psfMap}, aims to account for a possible point source contribution to the [\ion{C}{2}] flux. Here we assume a simple 2D Gaussian model for the point-spread function of the SOFIA telescope with $\mathrm{FWHM}\approx 12\farcs5$ at the observed [\ion{C}{2}] wavelength. In Fig.~\ref{fig:maps}, first and second panel, we show maps of the two components, both scaled to the observed map (fourth panel) by minimizing the least-square error between observation and model. It becomes apparent that the emission is extended and a point source is not a good representation. The scaling factors, when using only single components, are $a'=4.1$ for the extended emission and $b'=2.6$ for the point source model. 

\begin{figure}
\centering
\includegraphics[width=\linewidth]{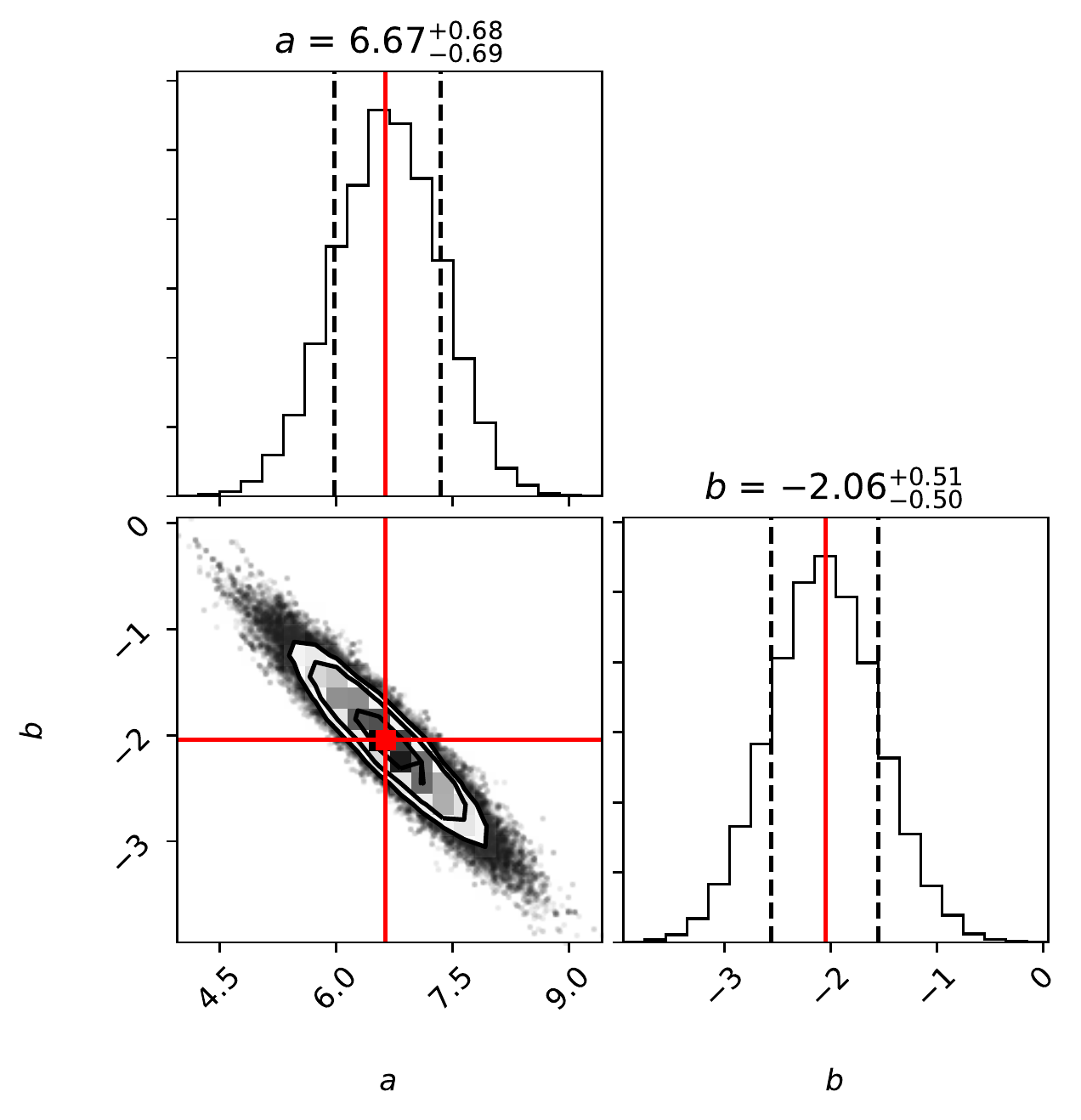}
\caption{Corner-plot visualization \citep{corner} of the parameter space explored by our model fitting procedure. $a$ and $b$ denote the fitted flux of the extended emission and the point source emission respectively. The histograms show the marginalized posterior distribution of the parameters, the dashed vertical lines indicate the 1$\sigma$ confidence intervals. Red lines show the maximum likelihood parameters.}
\label{fig:mcmc}
\end{figure}

The full model is given by a linear combination $a \times {\tt HaMap} + b \times {\tt psfMap}$. Since we normalize both reference maps to a total flux of 1, the scaling factors $a$ and $b$ denote the flux of the extended [\ion{C}{2}] emission and the point-source emission respectively, in units of $10^{-13}\,\mathrm{erg}\,\mathrm{s}^{-1}\,\mathrm{cm}^{-2}$. To find the best-fit parameters of this model, including their 1$\sigma$ uncertainties, we use the Markov-Chain Monte-Carlo code \textsc{Emcee} \citep{2013PASP..125..306F}. Figure \ref{fig:mcmc} is a visualization of the explored parameter space, the resulting map is shown in Fig.~\ref{fig:maps}, third panel. The best fit parameters are $a = 6.7\pm 0.7$ and $b = -2.1\pm 0.5$, which indicates that the extended emission is the dominating component and an additional small negative point source contribution is favored by the model. 

The total [\ion{C}{2}] flux indicated by the model is $4.6\times 10^{-13}\,\mathrm{erg}\,\mathrm{s}^{-1}\,\mathrm{cm}^{-2}$, the same as the flux derived from the Gaussian fit of the integrated spectrum. Comparing with the total extinction corrected H$\alpha$ flux of $5.6\times 10^{-13}\,\mathrm{erg}\,\mathrm{s}^{-1}\,\mathrm{cm}^{-2}$, we derive a flux ratio of $f_\mathrm{[CII]}/f_{\mathrm{H}\alpha}$$\sim$$0.8$. Assuming a linear $L_{[\mathrm{C II}]}-L_{\mathrm{H}\alpha}$ relation, we connect the linear $L_{[\mathrm{C II}]}-L_\mathrm{SFR}$ relation from \cite{2015ApJ...800....1H} with the linear $L_{\mathrm{H}\alpha}-\mathrm{SFR}$ relation \citep{2012ARA&A..50..531K} which leads to a predicted relation of $L_{[\mathrm{C II}]} = 1.1 \times L_{\mathrm{H}\alpha}$, which is consistent with our result.

The modeled total [\ion{C}{2}] flux corresponds to a [\ion{C}{2}] luminosity of $L_{[\mathrm{C II}]} = 3.5\times 10^8\,L_\odot$. Using the calibration of \citet{2015ApJ...800....1H}, who report a scatter of $\sim$$0.2\,\mathrm{dex}$, we obtain a star formation rate of $\sim 8.4\,M_\odot\,\mathrm{yr}^{-1}$ which agrees very well with the star formation estimate from the AGN-subtracted and dust-corrected H$\alpha$ luminosity of $\sim 8.9\,M_\odot\,\mathrm{yr}^{-1}$ \citep{2012ARA&A..50..531K}.

Since the magnitude $a=6.7\pm 0.7$ of the extended emission is significantly higher than the magnitude of the point source contribution $b = -2.1\pm 0.5$, we conclude that the extended emission is the dominating contribution to the [\ion{C}{2}] flux. However, as also visible in the maps in Fig.~\ref{fig:maps}, a point-like contribution is required. A negative point source contribution could indicate that the central [\ion{C}{2}] emission is reduced in the presence of an AGN, for example due to overionization of the emitting gas \citep[e.g.][]{2015A&A...580A...5L}. Also this could hint at different star formation rates in bar and bulge, of which the latter cannot be spatially separated from an AGN component at the resolution of SOFIA.

In Fig.~\ref{fig:maps}, \added{top row,} we show a map of the model together with maps of the two components separately. These maps are directly compared to the aforementioned map of the observed emission. 

\replaced{While the elongated bar-like structure becomes apparent in both, the observed [\ion{C}{2}] map as well as the model, there are considerable deviations between them. In particular, we observe additional [\ion{C}{2}] emission in the North. Since the emission region in the North of the map is outside of the bar and no star formation is visible in (extinction corrected) H$\alpha$, we suspect that the [\ion{C}{2}] emission might be tracing the cold neutral medium (CNM) of which [\ion{C}{2}] is the dominant cooling process \citep{2003ApJ...587..278W}.}
{In Fig.~\ref{fig:maps}, bottom row, we show surface brightnesses of star formation and [\ion{C}{2}] emission. At S/N$>3$ (indicated with crosses in the maps), we detect an elongated bar-like structure with an extend of almost one arcminute. Next to the maps, we place the pixels with S/N$>3$ in the $\Sigma_\mathrm{[CII]} - \Sigma_\mathrm{SFR}$ relation. It becomes apparent that in the central pixels the [\ion{C}{2}] emission follows the H$\alpha$ emission and lies on published $\Sigma_\mathrm{[CII]} - \Sigma_\mathrm{SFR}$ relations \citep{2014A&A...568A..62D,2015ApJ...800....1H}. Therefore, we can conclude that, within the galaxy, [\ion{C}{2}] predominantly traces star-formation excited PDRs. In the North and South of the galaxy, however, we observe additional [\ion{C}{2}] emission that is not following the H$\alpha$ and CO emission. We suspect that the [\ion{C}{2}] emission there is tracing the cold neutral medium (CNM) of which [\ion{C}{2}] is the dominant cooling process and which can be excited by many processes including the UV background and cosmic rays \citep[e.g.][]{2003ApJ...587..278W}. While in the centers of galaxies, PDRs are the main contributor to [\ion{C}{2}], in the outskirts of galaxies, the CNM can contribute significantly to the [\ion{C}{2}] emission \citep[see e.g. in M33;][]{2013A&A...553A.114K}. Consistent with this, the velocities of the emission in these regions are contained in the spectrum of \ion{H}{1} which is the main tracer of the CNM.}


\begin{figure}
\centering
\includegraphics[width=\columnwidth]{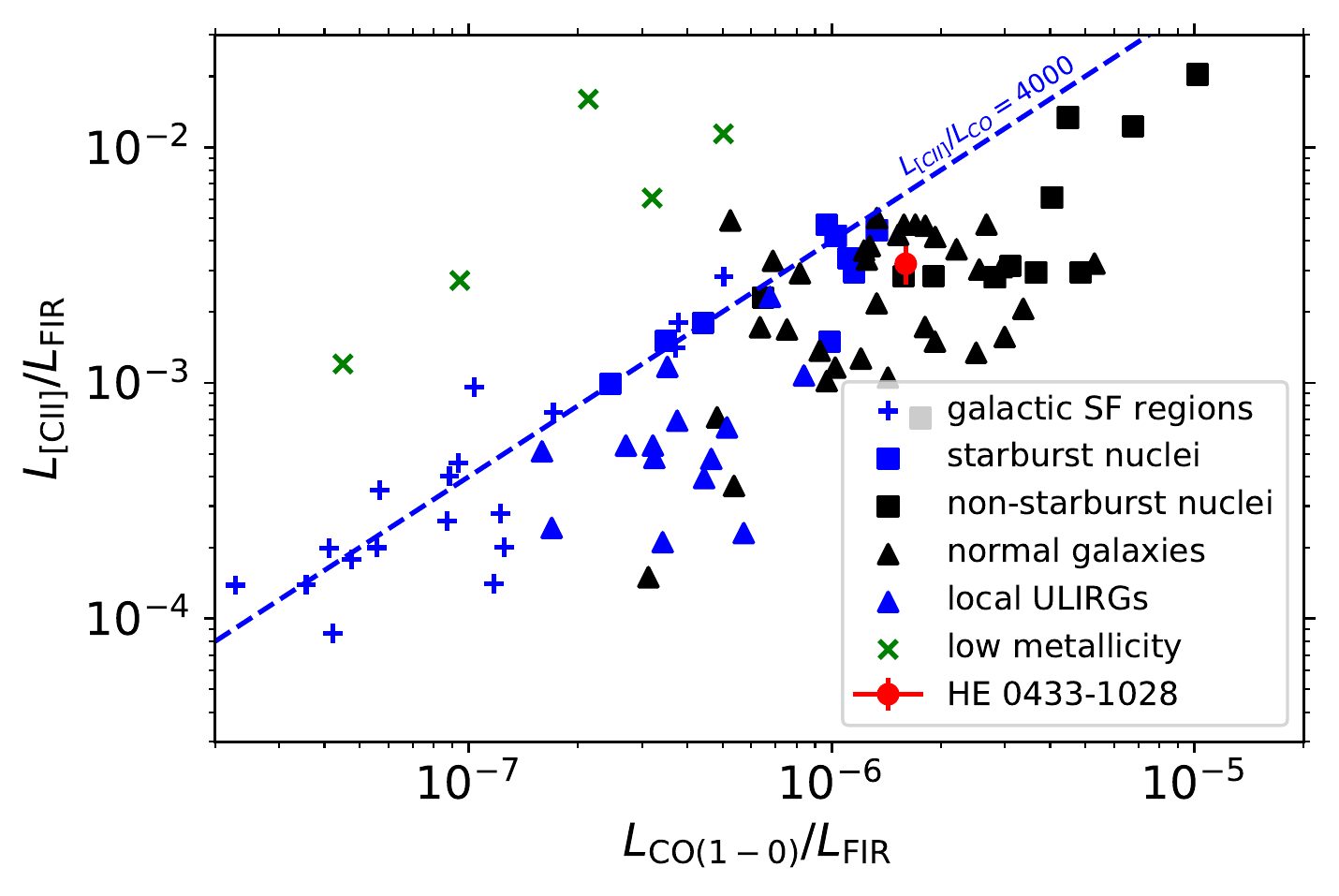}
\caption{Position of HE0433-1028 in the $L_\mathrm{CO(1-0)}/\mathrm{FIR} - L_{[\mathrm{C II}]}/L_\mathrm{FIR}$ plane compared to datapoints for normal galaxies, ULIRGs, galactic star formation regions, \replaced{and galactic nuclei}{galactic nuclei, and low-metallicity galaxies} from \citet{2010ApJ...724..957S} \added{and \citet{2011A&A...530L...8D}} (and references therein). A typical starburst [\ion{C}{2}]/CO line ratio of $\sim$$4000$ is indicated. \deleted{The green crosses denote low-metallicity galaxies collected by \cite{2011A&A...530L...8D}.}}
\label{fig:fir-plot}
\end{figure}

In Fig.~\ref{fig:fir-plot}, we show the position of HE0433-1028 in the $L_\mathrm{[CII]}/L_\mathrm{FIR} - L_{\mathrm{CO}(1-0)}/L_\mathrm{FIR}$ diagram \citep[][and references therein]{2010ApJ...724..957S}. The far-infrared luminosity is $L_\mathrm{FIR} = 1.1\times 10^{11}\,L_\odot$ \citep[derived in][]{2016A&A...587A.137M} which indicates that the [\ion{C}{2}] contributes around $0.5\%$ to the FIR luminosity. The integrated emission of the $^{12}$CO(1--0) transition, mapped with the Atacama Large Millimeter/submillimeter Array (ALMA, T. Davis, priv.~comm.), has a flux of $(58.05\pm 1.08)\,\mathrm{Jy}\,\mathrm{km}\,\mathrm{s}^{-1}$ which corresponds to a luminosity of $L_\mathrm{CO} = 1.74\times 10^5\,L_\odot$. This is consistent with previous single-dish CO-observations with the IRAM-30m telescope \citep{2007A&A...470..571B}, yielding a luminosity of the $^{12}$CO(1--0) line of $L_\mathrm{CO} = 1.1\times 10^5\,L_\odot$ which is $\sim$$30\%$ lower since the galaxy is more extended than the IRAM-30m beam.

The line ratio between the [\ion{C}{2}] and the $^{12}$CO(1--0) line is often used as a diagnostic of the star formation activity in galaxies. While the [\ion{C}{2}] line scales strongly with the far-UV radiation emitted by young OB stars, the CO(1--0) stems from better shielded regions. Therefore, starburst dominated regions show higher [\ion{C}{2}]/CO ratios. Typical ratios of starburst galaxies are of the order of $\sim 4000$ and can reach up to a few $10^4$ in extreme cases but are by a factor of $3$ smaller in non-starburst nuclei \citep{1991ApJ...373..423S}. 
Also, in low metallicity galaxies, \added{higher }$L_\mathrm{[CII]}/L_\mathrm{FIR}$ and $L_\mathrm{[CII]}/L_\mathrm{CO}$ ratios \deleted{several orders of magnitude higher} than in normal galaxies are observed \citep[e.g.][]{2000NewAR..44..249M,2006A&A...451..917R} \added{\citep{2015A&A...578A..53C}} which are explained by larger C$^+$ emitting regions.
For HE0433-1028, we measure a line ratio of $L_{[CII]}/L_\mathrm{CO} \sim 2030$ which is a typical value \deleted{and does not imply low metallicity}. A more detailed analysis of possible metallicity gradients will be pursued with spatially-resolved CO data. This underlines the necessity for multiwavelength observations with high spatial resolution.

\section{Summary and conclusions}
\label{sec:conclusion}

In this letter, we presented integral-field spectroscopy observations of the [\ion{C}{2}] $\lambda 158\,\mu \mathrm{m}$ emission line of the nearby luminous AGN HE 0433--1028 at a redshift of $0.0355$. This is the first spatially resolved far-infrared observation of a nearby luminous AGN and the object with the highest redshift that has been studied with the airborne observatory SOFIA so far.

Compared to high-z QSOs, HE 0433--1028 is still close enough to resolve separate star formation regions \citep{2015A&A...575A.128B}. This makes sources like HE 0433--1028 ideal targets to analyze a star forming galaxy in the presence of a strong AGN and study the counterplay of star formation and AGN activity.

We can clearly resolve the elongated structure of the star forming regions along the large-scale bar. We compare the data to a model which consists of an extended emission and point source component, both based on a high-resolution AGN-subtracted H$\alpha$-map. The model fit suggests that the [\ion{C}{2}] follows the extended ($30\arcsec \approx 20\,\mathrm{kpc}$) H$\alpha$ emission, which is excited by star formation. An additional small negative point source component is suggested by the model. However, a dependency of the point source contribution on the AGN luminosity or other parameters has to be verified with a larger sample.

Previous studies \citep[e.g.][]{2014ApJ...790...15S} already suggested that [\ion{C}{2}] can serve as a star formation estimator also in AGNs, but \replaced{with higher scatter}{only with a precision of $\sim 50\%$}. However, since they were comparing integrated measurements, it was not clear whether increased scatter comes from an additional component, associated with the AGN, or globally different scaling factors between [\ion{C}{2}] and SFR. With our spatially-resolved [\ion{C}{2}] observations and H$\alpha$ comparison data, we see that in HE0433-1028 the extended emission of [\ion{C}{2}] and H$\alpha$ are statistically matching and yield scaling factors consistent with inactive galaxies, with an additional flux excess North and South of the bar. While integrated [\ion{C}{2}]/CO line ratios indicate a metallicity typical for normal galaxies, possible metallicity variations will be analyzed in more detail with upcoming spatially-resolved optical and CO measurements.

While the sole purpose of this letter is to demonstrate the feasibility of spatially resolved analysis of [\ion{C}{2}] emission in nearby luminous AGNs, with reasonable observing time effort, the upcoming observations of a larger set of in total five QSOs with different AGN luminosities from the CARS survey will allow us to probe the [\ion{C}{2}]-SFR connection as a function of AGN luminosity and SFR.




\acknowledgments
Based in part on observations made with the NASA/DLR Stratospheric Observatory for Infrared Astronomy (SOFIA). SOFIA is jointly operated by the Universities Space Research Association, Inc. (USRA), under NASA contract NAS2-97001, and the Deutsches SOFIA Institut (DSI) under DLR contract 50 OK 0901 to the University of Stuttgart.
This work was supported by the Deutsche Forschungsgemeinschaft (DFG) via SFB 956, subprojects A2, and A6.
The work of S.B. and C.O. was supported by NSERC (Natural Sciences and Engineering Research Council of Canada).
M.G. is supported by NASA through Einstein Postdoctoral Fellowship Award Number PF5-160137 issued by the Chandra X-ray Observatory Center, which is operated by the SAO for and on behalf of NASA under contract NAS8-03060. Support for this work was also provided by Chandra grant GO7-18121X.
M.K. acknowledges support by DLR 50OR1802.
M.P.T. acknowledges support from the Spanish MINECO through grants AYA2012-38491-C02-02 and AYA2015-63939-C2-1-P.

\bibliographystyle{aasjournal}

\begin{thebibliography}{}
\expandafter\ifx\csname natexlab\endcsname\relax\def\natexlab#1{#1}\fi
\providecommand{\url}[1]{\href{#1}{#1}}

\bibitem[{{Bacon} {et~al.}(2010){Bacon}, {Accardo}, {Adjali}, {Anwand},
  {Bauer}, {Biswas}, {Blaizot}, {Boudon}, {Brau-Nogue}, {Brinchmann},
  {Caillier}, {Capoani}, {Carollo}, {Contini}, {Couderc}, {Daguis{\'e}},
  {Deiries}, {Delabre}, {Dreizler}, {Dubois}, {Dupieux}, {Dupuy}, {Emsellem},
  {Fechner}, {Fleischmann}, {Fran{\c c}ois}, {Gallou}, {Gharsa}, {Glindemann},
  {Gojak}, {Guiderdoni}, {Hansali}, {Hahn}, {Jarno}, {Kelz}, {Koehler},
  {Kosmalski}, {Laurent}, {Le Floch}, {Lilly}, {Lizon}, {Loupias}, {Manescau},
  {Monstein}, {Nicklas}, {Olaya}, {Pares}, {Pasquini}, {P{\'e}contal-Rousset},
  {Pell{\'o}}, {Petit}, {Popow}, {Reiss}, {Remillieux}, {Renault}, {Roth},
  {Rupprecht}, {Serre}, {Schaye}, {Soucail}, {Steinmetz}, {Streicher}, {Stuik},
  {Valentin}, {Vernet}, {Weilbacher}, {Wisotzki}, \&
  {Yerle}}]{2010SPIE.7735E..08B}
{Bacon}, R., {Accardo}, M., {Adjali}, L., {et~al.} 2010, in \procspie, Vol.
  7735, Ground-based and Airborne Instrumentation for Astronomy III, 773508

\bibitem[{{Baldwin} {et~al.}(1981){Baldwin}, {Phillips}, \&
  {Terlevich}}]{1981PASP...93....5B}
{Baldwin}, J.~A., {Phillips}, M.~M., \& {Terlevich}, R. 1981, \pasp, 93, 5

\bibitem[{{Bertram} {et~al.}(2007){Bertram}, {Eckart}, {Fischer}, {Zuther},
  {Straubmeier}, {Wisotzki}, \& {Krips}}]{2007A&A...470..571B}
{Bertram}, T., {Eckart}, A., {Fischer}, S., {et~al.} 2007, \aap, 470, 571

\bibitem[{{Boselli} {et~al.}(2002){Boselli}, {Gavazzi}, {Lequeux}, \&
  {Pierini}}]{2002A&A...385..454B}
{Boselli}, A., {Gavazzi}, G., {Lequeux}, J., \& {Pierini}, D. 2002, \aap, 385,
  454

\bibitem[{{Busch} {et~al.}(2015){Busch}, {Smaji{\'c}}, {Scharw{\"a}chter},
  {Eckart}, {Valencia-S.}, {Moser}, {Husemann}, {Krips}, \&
  {Zuther}}]{2015A&A...575A.128B}
{Busch}, G., {Smaji{\'c}}, S., {Scharw{\"a}chter}, J., {et~al.} 2015, \aap,
  575, A128

\bibitem[{{Busch} {et~al.}(2016){Busch}, {Fazeli}, {Eckart}, {Valencia-S.},
  {Smaji{\'c}}, {Moser}, {Scharw{\"a}chter}, {Dierkes}, \&
  {Fischer}}]{2016A&A...587A.138B}
{Busch}, G., {Fazeli}, N., {Eckart}, A., {et~al.} 2016, \aap, 587, A138

\bibitem[{{Cardelli} {et~al.}(1989){Cardelli}, {Clayton}, \&
  {Mathis}}]{1989ApJ...345..245C}
{Cardelli}, J.~A., {Clayton}, G.~C., \& {Mathis}, J.~S. 1989, \apj, 345, 245

\bibitem[{{Cid Fernandes} {et~al.}(2010){Cid Fernandes}, {Stasi{\'n}ska},
  {Schlickmann}, {Mateus}, {Vale Asari}, {Schoenell}, \&
  {Sodr{\'e}}}]{2010MNRAS.403.1036C}
{Cid Fernandes}, R., {Stasi{\'n}ska}, G., {Schlickmann}, M.~S., {et~al.} 2010,
  \mnras, 403, 1036

\bibitem[{{Cormier} {et~al.}(2015){Cormier}, {Madden}, {Lebouteiller}, {Abel},
  {Hony}, {Galliano}, {R{\'e}my-Ruyer}, {Bigiel}, {Baes}, {Boselli},
  {Chevance}, {Cooray}, {De Looze}, {Doublier}, {Galametz}, {Hughes},
  {Karczewski}, {Lee}, {Lu}, \& {Spinoglio}}]{2015A&A...578A..53C}
{Cormier}, D., {Madden}, S.~C., {Lebouteiller}, V., {et~al.} 2015, \aap, 578,
  A53

\bibitem[{{Davies} {et~al.}(2016){Davies}, {Groves}, {Kewley}, {Dopita},
  {Hampton}, {Shastri}, {Scharw{\"a}chter}, {Sutherland}, {Kharb}, {Bhatt},
  {Jin}, {Banfield}, {Zaw}, {James}, {Juneau}, \&
  {Srivastava}}]{2016MNRAS.462.1616D}
{Davies}, R.~L., {Groves}, B., {Kewley}, L.~J., {et~al.} 2016, \mnras, 462,
  1616

\bibitem[{{Davoust} \& {Contini}(2004)}]{2004A&A...416..515D}
{Davoust}, E., \& {Contini}, T. 2004, \aap, 416, 515

\bibitem[{{De Breuck} {et~al.}(2011){De Breuck}, {Maiolino}, {Caselli},
  {Coppin}, {Hailey-Dunsheath}, \& {Nagao}}]{2011A&A...530L...8D}
{De Breuck}, C., {Maiolino}, R., {Caselli}, P., {et~al.} 2011, \aap, 530, L8

\bibitem[{{De Looze} {et~al.}(2014){De Looze}, {Cormier}, {Lebouteiller},
  {Madden}, {Baes}, {Bendo}, {Boquien}, {Boselli}, {Clements}, {Cortese},
  {Cooray}, {Galametz}, {Galliano}, {Graci{\'a}-Carpio}, {Isaak}, {Karczewski},
  {Parkin}, {Pellegrini}, {R{\'e}my-Ruyer}, {Spinoglio}, {Smith}, \&
  {Sturm}}]{2014A&A...568A..62D}
{De Looze}, I., {Cormier}, D., {Lebouteiller}, V., {et~al.} 2014, \aap, 568,
  A62

\bibitem[{Fischer {et~al.}(2018)Fischer, Beckmann, Bryant, Colditz, Fumi, Geis,
  Hamidouche, Henning, H\"{o}nle, Iserlohe, Klein, Krabbe, Looney, Poglitsch,
  Raab, Rebell, Rosenthal, Savage, Schweitzer, Trinh, \& Vacca}]{Fischer2018}
Fischer, C., Beckmann, S., Bryant, A., {et~al.} 2018, Journal of Astronomical
  Instrumentation, 1840003.
\newblock \url{https://doi.org/10.1142/s2251171718400032}

\bibitem[{Foreman-Mackey(2016)}]{corner}
Foreman-Mackey, D. 2016, The Journal of Open Source Software, 24,
  doi:10.21105/joss.00024.
\newblock \url{http://dx.doi.org/10.5281/zenodo.45906}

\bibitem[{{Foreman-Mackey} {et~al.}(2013){Foreman-Mackey}, {Hogg}, {Lang}, \&
  {Goodman}}]{2013PASP..125..306F}
{Foreman-Mackey}, D., {Hogg}, D.~W., {Lang}, D., \& {Goodman}, J. 2013, \pasp,
  125, 306

\bibitem[{{Fruchter} \& {Hook}(2002)}]{2002PASP..114..144F}
{Fruchter}, A.~S., \& {Hook}, R.~N. 2002, \pasp, 114, 144

\bibitem[{{Graci{\'a}-Carpio} {et~al.}(2011){Graci{\'a}-Carpio}, {Sturm},
  {Hailey-Dunsheath}, {Fischer}, {Contursi}, {Poglitsch}, {Genzel},
  {Gonz{\'a}lez-Alfonso}, {Sternberg}, {Verma}, {Christopher}, {Davies},
  {Feuchtgruber}, {de Jong}, {Lutz}, \& {Tacconi}}]{2011ApJ...728L...7G}
{Graci{\'a}-Carpio}, J., {Sturm}, E., {Hailey-Dunsheath}, S., {et~al.} 2011,
  \apjl, 728, L7

\bibitem[{{Herrera-Camus} {et~al.}(2015){Herrera-Camus}, {Bolatto}, {Wolfire},
  {Smith}, {Croxall}, {Kennicutt}, {Calzetti}, {Helou}, {Walter}, {Leroy},
  {Draine}, {Brandl}, {Armus}, {Sandstrom}, {Dale}, {Aniano}, {Meidt},
  {Boquien}, {Hunt}, {Galametz}, {Tabatabaei}, {Murphy}, {Appleton}, {Roussel},
  {Engelbracht}, \& {Beirao}}]{2015ApJ...800....1H}
{Herrera-Camus}, R., {Bolatto}, A.~D., {Wolfire}, M.~G., {et~al.} 2015, \apj,
  800, 1

\bibitem[{{Herrera-Camus} {et~al.}(2018){Herrera-Camus}, {Sturm},
  {Graci{\'a}-Carpio}, {Lutz}, {Contursi}, {Veilleux}, {Fischer},
  {Gonz{\'a}lez-Alfonso}, {Poglitsch}, {Tacconi}, {Genzel}, {Maiolino},
  {Sternberg}, {Davies}, \& {Verma}}]{2018ApJ...861...95H}
{Herrera-Camus}, R., {Sturm}, E., {Graci{\'a}-Carpio}, J., {et~al.} 2018, \apj,
  861, 95

\bibitem[{{Hollenbach} \& {Tielens}(1999)}]{1999RvMP...71..173H}
{Hollenbach}, D.~J., \& {Tielens}, A.~G.~G.~M. 1999, Reviews of Modern Physics,
  71, 173

\bibitem[{{Husemann} {et~al.}(2014){Husemann}, {Jahnke}, {S{\'a}nchez},
  {Wisotzki}, {Nugroho}, {Kupko}, \& {Schramm}}]{2014MNRAS.443..755H}
{Husemann}, B., {Jahnke}, K., {S{\'a}nchez}, S.~F., {et~al.} 2014, \mnras, 443,
  755

\bibitem[{{Husemann} {et~al.}(2013){Husemann}, {Wisotzki}, {S{\'a}nchez}, \&
  {Jahnke}}]{2013A&A...549A..43H}
{Husemann}, B., {Wisotzki}, L., {S{\'a}nchez}, S.~F., \& {Jahnke}, K. 2013,
  \aap, 549, A43

\bibitem[{{Husemann} {et~al.}(2017){Husemann}, {Tremblay}, {Davis}, {Busch},
  {McElroy}, {Neumann}, {Urrutia}, {Krumpe}, {Scharw{\"a}chter}, {Powell},
  {Perez-Torres}, \& {The CARS Team}}]{2017Msngr.169...42H}
{Husemann}, B., {Tremblay}, G., {Davis}, T., {et~al.} 2017, The Messenger, 169,
  42

\bibitem[{{Kauffmann} {et~al.}(2003){Kauffmann}, {Heckman}, {Tremonti},
  {Brinchmann}, {Charlot}, {White}, {Ridgway}, {Brinkmann}, {Fukugita}, {Hall},
  {Ivezi{\'c}}, {Richards}, \& {Schneider}}]{2003MNRAS.346.1055K}
{Kauffmann}, G., {Heckman}, T.~M., {Tremonti}, C., {et~al.} 2003, \mnras, 346,
  1055

\bibitem[{{Kennicutt} \& {Evans}(2012)}]{2012ARA&A..50..531K}
{Kennicutt}, R.~C., \& {Evans}, N.~J. 2012, \araa, 50, 531

\bibitem[{{Kewley} {et~al.}(2001){Kewley}, {Dopita}, {Sutherland}, {Heisler},
  \& {Trevena}}]{2001ApJ...556..121K}
{Kewley}, L.~J., {Dopita}, M.~A., {Sutherland}, R.~S., {Heisler}, C.~A., \&
  {Trevena}, J. 2001, \apj, 556, 121

\bibitem[{{Klein} {et~al.}(2010){Klein}, {Poglitsch}, {Raab}, {Geis},
  {Hamidouche}, {Looney}, {H{\"o}nle}, {Nishikida}, {Genzel}, \&
  {Henning}}]{2010SPIE.7735E..1TK}
{Klein}, R., {Poglitsch}, A., {Raab}, W., {et~al.} 2010, in \procspie, Vol.
  7735, Ground-based and Airborne Instrumentation for Astronomy III, 77351T

\bibitem[{{Kramer} {et~al.}(2013){Kramer}, {Abreu-Vicente},
  {Garc{\'{\i}}a-Burillo}, {Rela{\~n}o}, {Aalto}, {Boquien}, {Braine},
  {Buchbender}, {Gratier}, {Israel}, {Nikola}, {R{\"o}llig}, {Verley}, {van der
  Werf}, \& {Xilouris}}]{2013A&A...553A.114K}
{Kramer}, C., {Abreu-Vicente}, J., {Garc{\'{\i}}a-Burillo}, S., {et~al.} 2013,
  \aap, 553, A114

\bibitem[{{Langer} \& {Pineda}(2015)}]{2015A&A...580A...5L}
{Langer}, W.~D., \& {Pineda}, J.~L. 2015, \aap, 580, A5

\bibitem[{{Lord}(1992)}]{1992nstc.rept.....L}
{Lord}, S.~D. 1992, {A new software tool for computing Earth's atmospheric
  transmission of near- and far-infrared radiation}, Tech. rep.

\bibitem[{{Luhman} {et~al.}(2003){Luhman}, {Satyapal}, {Fischer}, {Wolfire},
  {Sturm}, {Dudley}, {Lutz}, \& {Genzel}}]{2003ApJ...594..758L}
{Luhman}, M.~L., {Satyapal}, S., {Fischer}, J., {et~al.} 2003, \apj, 594, 758

\bibitem[{{Madden}(2000)}]{2000NewAR..44..249M}
{Madden}, S.~C. 2000, \nar, 44, 249

\bibitem[{{Moser} {et~al.}(2016){Moser}, {Krips}, {Busch}, {Scharw{\"a}chter},
  {K{\"o}nig}, {Eckart}, {Smaji{\'c}}, {Garc{\'{\i}}a-Marin}, {Valencia-S.},
  {Fischer}, \& {Dierkes}}]{2016A&A...587A.137M}
{Moser}, L., {Krips}, M., {Busch}, G., {et~al.} 2016, \aap, 587, A137

\bibitem[{{R{\"o}llig} {et~al.}(2006){R{\"o}llig}, {Ossenkopf}, {Jeyakumar},
  {Stutzki}, \& {Sternberg}}]{2006A&A...451..917R}
{R{\"o}llig}, M., {Ossenkopf}, V., {Jeyakumar}, S., {Stutzki}, J., \&
  {Sternberg}, A. 2006, \aap, 451, 917

\bibitem[{{Sargsyan} {et~al.}(2014){Sargsyan}, {Samsonyan}, {Lebouteiller},
  {Weedman}, {Barry}, {Bernard-Salas}, {Houck}, \&
  {Spoon}}]{2014ApJ...790...15S}
{Sargsyan}, L., {Samsonyan}, A., {Lebouteiller}, V., {et~al.} 2014, \apj, 790,
  15

\bibitem[{{Scharw{\"a}chter} {et~al.}(2011){Scharw{\"a}chter}, {Dopita},
  {Zuther}, {Fischer}, {Komossa}, \& {Eckart}}]{2011AJ....142...43S}
{Scharw{\"a}chter}, J., {Dopita}, M.~A., {Zuther}, J., {et~al.} 2011, \aj, 142,
  43

\bibitem[{{Smith} {et~al.}(2017){Smith}, {Croxall}, {Draine}, {De Looze},
  {Sandstrom}, {Armus}, {Beir{\~a}o}, {Bolatto}, {Boquien}, {Brandl},
  {Crocker}, {Dale}, {Galametz}, {Groves}, {Helou}, {Herrera-Camus}, {Hunt},
  {Kennicutt}, {Walter}, \& {Wolfire}}]{2017ApJ...834....5S}
{Smith}, J.~D.~T., {Croxall}, K., {Draine}, B., {et~al.} 2017, \apj, 834, 5

\bibitem[{{Stacey} {et~al.}(1991){Stacey}, {Geis}, {Genzel}, {Lugten},
  {Poglitsch}, {Sternberg}, \& {Townes}}]{1991ApJ...373..423S}
{Stacey}, G.~J., {Geis}, N., {Genzel}, R., {et~al.} 1991, \apj, 373, 423

\bibitem[{{Stacey} {et~al.}(2010){Stacey}, {Hailey-Dunsheath}, {Ferkinhoff},
  {Nikola}, {Parshley}, {Benford}, {Staguhn}, \&
  {Fiolet}}]{2010ApJ...724..957S}
{Stacey}, G.~J., {Hailey-Dunsheath}, S., {Ferkinhoff}, C., {et~al.} 2010, \apj,
  724, 957

\bibitem[{{Weilbacher} {et~al.}(2012){Weilbacher}, {Streicher}, {Urrutia},
  {Jarno}, {P{\'e}contal-Rousset}, {Bacon}, \&
  {B{\"o}hm}}]{2012SPIE.8451E..0BW}
{Weilbacher}, P.~M., {Streicher}, O., {Urrutia}, T., {et~al.} 2012, in
  \procspie, Vol. 8451, Software and Cyberinfrastructure for Astronomy II,
  84510B

\bibitem[{{Wolfire} {et~al.}(2003){Wolfire}, {McKee}, {Hollenbach}, \&
  {Tielens}}]{2003ApJ...587..278W}
{Wolfire}, M.~G., {McKee}, C.~F., {Hollenbach}, D., \& {Tielens}, A.~G.~G.~M.
  2003, \apj, 587, 278

\bibitem[{{Young} {et~al.}(2012){Young}, {Becklin}, {Marcum}, {Roellig}, {De
  Buizer}, {Herter}, {G{\"u}sten}, {Dunham}, {Temi}, {Andersson}, {Backman},
  {Burgdorf}, {Caroff}, {Casey}, {Davidson}, {Erickson}, {Gehrz}, {Harper},
  {Harvey}, {Helton}, {Horner}, {Howard}, {Klein}, {Krabbe}, {McLean}, {Meyer},
  {Miles}, {Morris}, {Reach}, {Rho}, {Richter}, {Roeser}, {Sandell}, {Sankrit},
  {Savage}, {Smith}, {Shuping}, {Vacca}, {Vaillancourt}, {Wolf}, \&
  {Zinnecker}}]{2012ApJ...749L..17Y}
{Young}, E.~T., {Becklin}, E.~E., {Marcum}, P.~M., {et~al.} 2012, \apjl, 749,
  L17

\end{thebibliography}



\end{document}